# On the weak nematic elasticity and soft deformation modes

A. I. Leonov[1a] and V.S. Volkov[b]

[a)] *Department of Polymer Engineering, The University of Akron, Akron, OH 44325-0301, USA*
[b)] *Laboratory of Rheology, Institute of Petrochemical Synthesis, Russian Academy of Sciences, Leninsky Pr., 29, Moscow 117912 Russia.*

**Abstract**

The paper investigates the general case of incompressible non-classical elasticity with small deformations and rotations. It is shown that the internal spin rotations are important only when a specific Born term is presented in the free energy. The thermodynamic stability conditions and existence of soft deformational modes are analysed for free energy with three rotational degrees of freedom and scalar order parameter. It is shown that the soft deformation modes do not always exist. If the conditions for existence of soft modes are satisfied and the Born term is absent, the stress tensor is proved to be symmetric. It is shown that the infinitesimal Warner potential satisfies these conditions. When the stress symmetry could be assumed only approximately, a simplified, "reduced" relations are derived for the free energy and stress. The reduce formulation, which preserves the general stability conditions and the condition for existence of soft deformations, also allows easy calculations of stress-strain-rotation relations.

## 1. Introduction

Liquid crystalline (LC), cross-linked elastomers are solids that contain some rigid molecular elements such as mesogenic groups inserted in or the side branches attached to the main (backbone) polymer chains. An important feature of LC elastomers is that they display an additional degree of freedom, internal rotation. Incorporating this effect in a continuum approach allows understanding some unusual deformational phenomena for these anisotropic solids.

In order to describe macroscopic elasticity of LC elastomers in the equilibrium (static) limit several phenomenological theories have been developed, which included internal rotations in elastic free energy. Using symmetry arguments, De Gennes (1980) first introduced the internal rotations in the free energy density for nematic elastomers (see also Halperin (1986)). From formal viewpoint, de Gennes (1980) phenomenology is reminiscent the Ericksen (1960) - Leslie (1968) approach for low molecular weight LC nematic liquids. These theories involved into consideration only two rotational degrees of freedom, neglecting spin, i.e. the internal rotations around the axis of one-dimensional isotropy. Brand (1989) analysed effects of electromagnetic fields in rubber nematic elasticity. Terentjev (1993) applied a group-theoretical approach to analyse the linear elasticity and piezoelectric effects in these solids. Brand & Pleiner (1994) and Weilepp & Brand (1996) have also studied the rubber nematodynamics at the continuum level (see also Anderson et al (1999), Terentjev & Warner (2001)).

---

[1] Corresponding author.
 Tel.: +1-330-972-5138;  fax: +1-330-258-2339
  *E-mail address:* leonov@uakron.edu (A.I. Leonov).

To derive the free energy expression for nematic elastomers the classic theory of rubber elasticity was generalized on the basis of simple models of anisotropic fluctuations in anisotropic environment (e.g. see Abramchuk & Khokhlov (1987), Warner et al (1988), Warner & Wang (1991), Bladon et al (1993), Warner (1999)). When nematic interactions are presented, the chain free energy, even in the Gaussian phantom network approach, has an energetic component, which is a function of the degree of alignment. The first three molecular theories cited above were based on presentation of elastic free energy for deformed molecular network in terms of principal stretches. Bladon et al (1993) derived a general tensor expression for the energy of rubber-like nematic solids. The above molecular models analysed only the case with uniformly aligned rigid elastomeric groups oriented in a uniform external field. Terentjev et al (1996) developed a molecular theory of nematic rubber elasticity in the presence of non-uniform director field. The cited above molecular models described the effect of soft elasticity in nematic elastomers, unusual for conventional elasticity. Warner and co-workers (1994) were first who theoretically described the "soft deformation modes" that cost no elastic energy when accompanied by director rotation (e.g. see Olmsted (1994) and Warner (1999)). Theoretical predictions of these striking effects within a continuum phenomenology, confirmed later experimentally, is a great success of physicists contributed highly in this new field of continuum mechanics (e.g. see recent review by Lubensky et al (2002)).

Although this successful phenomenological theory was mainly inspired by particular molecular models, developing a general continuum approach, which is free of particular assumptions of specific molecular models, is still desirable. Along with searching for new possible behaviour for new types of nematic elastomers, there is a high need for developing a general theory applicable to many new types of materials with possible nematic behavior, such as soft nanocomposites and soft tissues, which may not belong to nematic elastomers. Such a theory still does not exist today.

From theoretical viewpoint, developing general theory is also required for resolving several problems and answering several questions. The most important is establishing the general conditions for the soft modes to exist. Until now, these conditions have been analysed only for particular (mostly Warner type) constitutive theories, which might create a misperception of the generality of those results. The possible effect of stresses/deformations on the spin of internal rotation and the scalar order parameter is another problem of interests. Another problem is about possible negligence of asymmetric part of stress tensor. Developing a reduced approach with (sometimes justified) stress symmetry assumption highly simplifies the general theory while still capturing its most basic features. It is especially important for formulation of a general nonlinear theory with finite deformations and internal rotations, as well as for developing much more complicated non-equilibrium theories for solid and liquid-like types of nematic viscoelasticity.

To answer the above questions we developed a general continuum theory of weak elasticity for nematic solids with three rotational degrees of freedom that describes a general anisotropic elastic behavior for incompressible nematic solids at small deformations and rotations. We secondly present a complete analysis of the "soft" deformation modes in nematic solids. To reveal these modes we apply to general situation a direct method, which has been employed before in several papers (e.g. see Warner (1999)). We thirdly develop a "reduced" formulation of the general theory, which is easy to use in non-equilibrium case. Finally, we demonstrate some new results obtained for linear version of the Warner potential.

Keeping in mind that general approach to nematic elasticity, even in linear limit, is important for developing non-equilibrium (relaxation) theories for solids and liquids (e.g. see Leonov and Volkov (2002)), we also tried to indicate in the following text the possible non-equilibrium effects.

**2. Reorientations in nematic solids**

From the macroscopic viewpoint, LC elastomers can be considered as anisotropic solids with three (internal) rotational degrees of freedom. Two of them are related to the change in orientation (or reorientation) of director, a unit vector $\underline{n}$, and the third one to rotation about the director,. Orientation degree in nematics is commonly described by the second rank tensor $\underline{\underline{Q}}$ defined as:

$$\underline{\underline{Q}} = s(\underline{n}\underline{n} - \underline{\underline{\delta}}/3).$$

Here $\underline{\underline{\delta}}$ is the unit tensor, and $s$ is the scalar order parameter, characterizing the degree of order. The existing theories of nematic elastomers usually assume that direction of preferred orientation changes only due to some external actions. In this case, parameter $s$ can be treated as a constant. For the rigid rotations of director, its rate of change $\underline{\dot{n}}$ is expressed as:

$$\underline{\dot{n}} = \underline{\omega}^I \times \underline{n} = -\underline{\underline{\omega}}^I \cdot \underline{n}, \text{ or } \dot{n}_i = \delta_{ijk}\omega^I_j n_k = -\omega^I_{ik} n_k. \quad (1)$$

Here $\underline{\omega}^I$ and $\underline{\underline{\omega}}^I$ are the vector and respective tensor, characterizing the speed of internal rotation, and hereafter overdot denotes the time derivative.

For elastic solids of nematic type, an algebraic kinematical relation between the initial value of director $\underline{n}_0$ in non-deformed state and its actual value $\underline{n}$ in the deformed state, equivalently substitutes the rate equation (1). This algebraic relation for reorientation of director written via an orthogonal tensor $\underline{\underline{R}}$ is:

$$\underline{n} = \underline{\underline{R}} \cdot \underline{n}_0; \quad \underline{\underline{R}} = \exp(-\underline{\underline{\Omega}}^I). \quad (2)$$

Here $\underline{\underline{\Omega}}^I$ is the anti-symmetric tensor of finite internal rotations. Introducing the vector of internal rotations $\underline{\Omega}^I$ as

$$\Omega^I_{ik} = -\delta_{ijk}\Omega^I_j, \quad (3)$$

represents (2) in the equivalent form:

$$\underline{n} - \underline{n}_0 = [\exp(\underline{\Omega}^I \times)]\underline{n}_0 \equiv \left(\sum_{k=1}^{\infty} \frac{1}{k!}(\underline{\Omega}^I \times)^k\right)\underline{n}_0 \quad (4)$$

Reorientation equation (4) shows that $\underline{n} = \underline{n}_0$, when $\underline{\Omega}^I = \underline{\Omega}^I_\| \equiv \lambda \underline{n}_0$.

In general case, one can decompose the vector (or corresponding tensor) of internal rotation $\underline{\Omega}^I$ in the sum,

$$\underline{\Omega}^I = \underline{\Omega}^I_\| + \underline{\Omega}^I_\perp, \quad (5)$$

where $\underline{\Omega}_{\parallel}^I$ and $\underline{\Omega}_{\perp}^I$ are respectively the vector-components parallel and orthogonal to the initial director $\underline{n}_0$. Nevertheless in spite of (5), the internal spin of finite rotations $\underline{\Omega}_{\parallel}^I$ does not disappear from the equivalent nonlinear kinematical relations (2) or (4).

In the linear case when the internal rotations are small enough, the first relation in (2) takes the approximate form:

$$\underline{n} - \underline{n}_0 \approx -\underline{\underline{\Omega}}^I \cdot \underline{n}_0 = \underline{\Omega}^I \times \underline{n}_0; \quad \left|\underline{\underline{\Omega}}^I\right| = \left|\underline{\Omega}^I\right| << 1. \qquad (6)$$

In this (and only in this) case, the linear equation (6) can be represented in the equivalent form:

$$\underline{n} - \underline{n}_0 = \underline{\Omega}_{\perp}^I \times \underline{n}_0, \quad \text{or} \quad \underline{\Omega}_{\perp}^I = -\underline{n} \times \underline{n}_0. \qquad (7)$$

This equation demonstrates that small reorientations of director are independent of its spin.

When the orthogonal tensor $\underline{\underline{R}}$ is time dependent, the well-known relation, $\underline{\dot{n}} = \underline{\underline{\dot{R}}} \cdot \underline{n}_0 = \underline{\underline{\dot{R}}} \cdot \underline{\underline{R}}^{-1} \cdot \underline{n}$, can be obtained by differentiating the first relation (2) with respect to time. Comparing this relation with (1) yields:

$$\underline{\underline{\omega}}^I = -\underline{\underline{\dot{R}}} \cdot \underline{\underline{R}}^{-1}. \qquad (8)$$

The formulae,

$$\underline{\underline{\omega}}^I \approx \underline{\underline{\dot{\Omega}}}^I, \quad \underline{\omega}^I = \underline{\dot{\Omega}}^I. \qquad (9)$$

are evidently valid for the linear case.

## 3. Dynamic equations

E.&F. Cosserat (1909) initiated first the continuum theory of anisotropic solids with internal rotations. This theory treats the orientation of each particle in continuum as independent of its position. Oseen (1925) developed later a similar theory for liquid crystals. In these theories, along with the common momentum balance equation

$$\rho \underline{\dot{v}} = \underline{\nabla} \cdot \underline{\underline{\sigma}} + \rho \underline{f}, \qquad (10)$$

the rotational degrees of freedom are also described by the dynamic equation for internal angular velocity,

$$\rho \underline{\dot{L}} = \underline{\nabla} \cdot \underline{\underline{\mu}} + \underline{\sigma}^a + \rho \underline{l}, \qquad \sigma_i^a = \delta_{ijk} \sigma_{jk}. \qquad (11)$$

In (10) and (11), $\rho$ is the density, $\underline{v}$ is the velocity, $\underline{\underline{\sigma}}$ is generally non-symmetric stress tensor, with anti-symmetric part $\underline{\underline{\sigma}}^a$, $\underline{f}$ is the body force, $\underline{L} = \underline{\underline{I}} \cdot \underline{\omega}^I$ is the internal angular moment, $\underline{\omega}^I$ is the internal angular velocity, $\underline{\underline{I}}$ is the rotational inertia tensor, $\underline{\underline{\mu}}$ is the couple stress tensor, and $\underline{l}$ is the body couple. Thus the stressed state of a continuum with internal rotations is characterized by generally non-symmetric tensors $\underline{\underline{\sigma}}$ and $\underline{\underline{\mu}}$. E.&F. Cosserat (1909) obtained the first equation in (11) using a variational method. In the Cosserat model with free rotation, internal angular velocity $\underline{\omega}^I$ is considered as an independent variable, consisting of two parts, the rotational velocity of the symmetry axis and spin, i.e. the rotational velocity of continuum about the symmetry axis.

The alternative form of the couple-stress equation (11) is:

$$\rho \underline{\underline{\dot{S}}} = 2\underline{\underline{\sigma}}^a + \underline{\underline{m}} + \underline{l}, \quad S_{ij} = \delta_{ijk} L_k, \tag{12}$$

where

$$m_{ij} = \delta_{ijk} \nabla_e \mu_{ek}. \tag{13}$$

Here $\underline{\underline{S}}$ is the internal moment of momentum and $\underline{l}$ is the body couple. Equation (12) is the balance equation for the internal angular momentum. De Groot and Mazur (1962) who also considered such an equation omitted the angular-momentum flux $F_{ije} = \delta_{ijk} \mu_{ek}$. In the non-polar case, when internal angular momentum, couple stresses, and body moments are negligible, the stress tensor according to (12) is symmetric.

## 4. Constitutive equations for incompressible weakly elastic nematics

We derive below general constitutive equations for weakly elastic nematic solids, when the elastic strain tensor $\underline{\underline{E}}$, the tensors of internal $\underline{\underline{\Omega}}^I$ and total (body) $\underline{\underline{\Omega}}$ rotations are assumed to be small. Here the tensors $\underline{\underline{E}}$ and $\underline{\underline{\Omega}}$ are defined via the displacement vector $\underline{u}$ by the common formulae of linear elasticity,

$$2\underline{\underline{E}} = \nabla \underline{u} + (\nabla \underline{u})^T; \quad 2\underline{\underline{\Omega}} = \nabla \underline{u} - (\nabla \underline{u})^T. \tag{14}$$

To simplify the analysis we consider below the incompressible case when $tr\underline{\underline{E}} = 0$. As independent state variables we introduce a scalar parameter $s_r \equiv s - s_0$, characterizing the difference between the values $s$ and $s_0$ of scalar order parameter in deformed and non-deformed states, and two frame invariant tensors, $\underline{\underline{E}}$ and $\underline{\underline{\Omega}}^r$ ($\equiv \underline{\underline{\Omega}} - \underline{\underline{\Omega}}^I$). If the state variables $\underline{\underline{E}}$ and $\underline{\underline{\Omega}}^r$ are known, the total internal and body rotations are known separately and the director in the deformed state is found from (6). Additionally, the initial value of director, $\underline{n}_0$, which characterizes the uniaxial anisotropy in non-deformed state, is considered in this theory as given.

Following Olmsted (1994) and Warner (1999), we will further neglect the dependence of the free energy on the director gradient, and consider in the following only the "deformational" part of free energy. Neglecting the director gradients (i.e. Frank's (1958) elasticity), can be justified for nematic elastomers within the length scales more than $\sqrt{k/\mu} \approx 10^{-3} cm$. Here $k$ ($\sim 10^{-5} N$) and $\mu$ ($\sim 10^5 Pa$) are the Frank and rubber elastic moduli, respectively.

A general relation for the Helmholtz free energy searched within this approach should be invariant relative to $\underline{n}_0 \to -\underline{n}_0$ transformation and quadratic in the state variables. Its general form for the linear case is:

$$\rho F = (G_s/2)s_r^2 + G_{sa}s_r tr(\underline{n}_0\underline{n}_0 \cdot \underline{\underline{E}}) + (G_0/2)tr\underline{\underline{E}}^2 + G_1 tr(\underline{n}_0\underline{n}_0 \cdot \underline{\underline{E}}^2) + G_2 tr^2(\underline{n}_0\underline{n}_0 \cdot \underline{\underline{E}})$$
$$- 2G_3 tr(\underline{n}_0\underline{n}_0 \cdot \underline{\underline{E}} \cdot \underline{\underline{\Omega}}^r) - G_4 tr[\underline{n}_0\underline{n}_0 \cdot (\underline{\underline{\Omega}}^r)^2] - (G_5/2)tr(\underline{\underline{\Omega}}^r)^2 \quad (tr\underline{\underline{E}}=0) \quad (15)$$

Equation (15) is an extended version of equations proposed by de Gennes (1980) and Olmsted (1994) where the Frank contributions are neglected. De Gennes and Olmsted equations written in vector form, are presented in our notations as:

$$2\rho F = \mu_0 (\underline{n}_0 \cdot \underline{\underline{E}} \cdot \underline{n}_0)^2 + \mu_1 (\underline{n}_0 \times \underline{\underline{E}} \times \underline{n}_0)^2 + \mu_2 (\underline{n}_0 \cdot \underline{\underline{E}} \times \underline{n}_0)^2 +$$
$$+ \alpha_1 (\underline{\underline{\Omega}}^r \times \underline{n}_0)^2 + \alpha_2 \underline{n}_0 \cdot \underline{\underline{E}} \cdot \underline{\underline{\Omega}}^r \times \underline{n}_0 \quad (16)$$

The following formulae,

$$\underline{n}_0 \times \underline{\underline{E}} \times \underline{n}_0 = \delta_{ejk} n_{0j} E_{ke} n_{0m} \delta_{mes} \quad \text{and} \quad \underline{n}_0 \cdot \underline{\underline{E}} \times \underline{n}_0 = n_{0k} E_{ke} n_{0m} \delta_{mes},$$

make possible to express (16) in the form (15) where

$$G_0 = \mu_1, \; G_1 = \mu_2/2 - \mu_1, \; G_2 = (\mu_0 + \mu_1 - \mu_2)/2, \; G_3 = \alpha_2/4, \; G_4 = \alpha_1/2,$$
$$G_s = G_{sa} = G_5 = 0. \quad (17)$$

Three new terms are involved in (15). The first two there describe the effects of deformation on scalar order parameter. The last term, $\sim G_5$, is term similar to that first proposed by Born (1920) for describing the internal rotations effects in isotropic fluids. This term is not of nematic nature, because it does not vanish in isotropic state. Because for isotropic elastic solids and liquids the Born effects are usually negligible, one can assume that the value of the "modulus" $G_5$ is considerably smaller than values of $G_0, G_1, G_2, G_3$ and $G_4$. Nevertheless, we will show that the Born term may cause several effects in weak nematic elasticity. The thermodynamic stability conditions imposed on material parameters in (15) are analyzed in the next Section.

The symmetric, $\underline{\underline{\sigma}}^s$, and anti-symmetric, $\underline{\underline{\sigma}}^a$ parts of "extra" stress tensor are calculated using free energy (15) as follows:

$$\underline{\underline{\sigma}}^s = \rho \partial F / \partial \underline{\underline{E}} = G_{sa} s_r \underline{n}_0 \underline{n}_0 + G_0 \underline{\underline{E}} + G_1(\underline{n}_0 \underline{n}_0 \cdot \underline{\underline{E}} + \underline{\underline{E}} \cdot \underline{n}_0 \underline{n}_0) + 2G_2 \underline{n}_0 \underline{n}_0 tr(\underline{\underline{E}} \cdot \underline{n}_0 \underline{n}_0)$$
$$+ G_3(\underline{n}_0 \underline{n}_0 \cdot \underline{\underline{\Omega}}^r - \underline{\underline{\Omega}}^r \cdot \underline{n}_0 \underline{n}_0) \quad (18a)$$

$$\underline{\underline{\sigma}}^a = \rho \partial F / \partial \underline{\underline{\Omega}} = G_3(\underline{n}_0 \underline{n}_0 \cdot \underline{\underline{E}} - \underline{\underline{E}} \cdot \underline{n}_0 \underline{n}_0) + G_4(\underline{n}_0 \underline{n}_0 \cdot \underline{\underline{\Omega}}^r + \underline{\underline{\Omega}}^r \cdot \underline{n}_0 \underline{n}_0) + G_5 \underline{\underline{\Omega}}^r . \quad (18b)$$

Note that the total stress tensor is defined as: $\underline{\underline{\sigma}} = -p\underline{\underline{\delta}} + \underline{\underline{\sigma}}^s + \underline{\underline{\sigma}}^a$. Here $p$ is an isotropic pressure.

Another equilibrium condition, $\rho \partial F / \partial s = 0$, for small variations of the scalar order parameter $s_r$, allows to express $s_r$ via the strain tensor $\underline{\underline{E}}$ as:

$$s_r = -(G_{sa} / G_s) tr(\underline{\underline{E}} \cdot \underline{n}_0 \underline{n}_0) . \quad (18c)$$

It is easy to see that inserting (18c) into (15) and (18a), and using instead of $G_2$, another parameter,

$$\hat{G}_2 = G_2 - 1/2 G_{sa}^2 / G_s , \quad (19)$$

allows us formally excluding the scalar parameter $s_r$ from consideration. However, in non-equilibrium situation, the variation of scalar order parameter would produce an additional kinetic equation.

Note that in all the equations of this Section, one can substitute the initial value of director $\underline{n}_0$ for the actual one $\underline{n}$ within the same precision. It means that the the free energy formulations by de Gennes (1980) and Olmsted (1994) where $\underline{n}$ and $\underline{n}_0$ were respectively employed, are equivalent. Then the reciprocal relation,

$$\underline{n}_0 - \underline{n} \approx \underline{\underline{\Omega}}^I \cdot \underline{n} ; \quad |\underline{\underline{\Omega}}^I| << 1 . \quad (6a)$$

might be useful due to (8). In such a description, the free energy $F$ in (15) will now depend on the scalar parameter $s_r$, strain $\underline{\underline{E}}$, relative rotation tensor $\underline{\underline{\Omega}}^r$, and on the actual value of director $\underline{n}$, which is now also variable. Since the director is a unit vector, another Lagrange multiplier $q$ should be introduced when allowing variations of $\underline{n}$. Thus instead of free energy expression $F(s_r, \underline{n}, \underline{\underline{E}}, \underline{\underline{\Omega}}^r)$, shown in (15) with $\underline{n}_0$ changes to $\underline{n}$, the modified free energy function, $\rho\tilde{F} = \rho F(s_r, \underline{n}, \underline{\underline{E}}, \underline{\underline{\Omega}}^r) - q\underline{n}\cdot\underline{n}$, should be introduced. Then an equilibrium condition, additional to (19), where $\underline{n}_0$ changes to $\underline{n}$, is of the form: $\rho\partial F/\partial\underline{n} = 2q\underline{n}$. The scalar multiplication this equation by $\underline{n}$ results in: $q = 1/2\rho\underline{n}\cdot\partial F/\partial\underline{n}$. This expression shows that the Lagrange multiplier $q$ is a quadratic order of magnitude, i.e. it is of the order of magnitude smaller than the stresses and the variation of the order parameter. Although this approach does not contain new physical information in equilibrium, it might be useful for developing non-equilibrium, relaxation models where the algebraic equation (6a) will be changed for an evolution equation for $\underline{n}$.

We now consider the effects of the above constitutive equations on the dynamics of internal rotations. In the case of weak nematic solids under study the inertia tensor $\underline{\underline{I}}$ is represented as:

$$\underline{\underline{I}} = I_\perp \underline{\underline{\delta}} + (I_\parallel - I_\perp)\underline{n}_0\underline{n}_0 \ . \tag{20}$$

Here $I_\perp$ and $I_\parallel$ are the principal values of the inertia tensor, the spin inertia being characterized by $I_\parallel$. When assuming negligible the effects of orientation gradients and the body moment density field, multiplying (11) and (12) by $\underline{n}_0$ and using (7) and (20) yields

$$\rho I_\perp \underline{\ddot{n}} = 2\underline{n}_0 \cdot \underline{\underline{\sigma}}^a , \qquad \rho I_\parallel \ddot{\Omega}_\parallel^I = \delta_{ijk} n_{i0} \underline{\sigma}_{jk}^a \quad (\Omega_\parallel^I = \underline{\Omega}^I \cdot \underline{n}_0). \tag{21}$$

Substituting now (7) and (18b) into (21) results in the dynamic equations for orientation,

$$\rho I_\perp \underline{\ddot{n}} = (G_5 + G_4)\underline{n}_0 \cdot \underline{\underline{\Omega}}^r + G_3[\underline{n}_0 \cdot \underline{\underline{E}} - \underline{n}_0 tr(\underline{\underline{E}} \cdot \underline{n}_0\underline{n}_0)], \tag{22a}$$

and spin,

$$\rho I_\parallel \ddot{\Omega}_\parallel^I = 2G_5(\Omega_\parallel - \Omega_\parallel^I) \quad (\Omega_\parallel = \underline{\Omega} \cdot \underline{n}_0). \tag{22b}$$

Equation (22b) clearly shows the importance of the Born term ($\sim G_5$). It demonstrates that $\Omega_\parallel \neq 0$ whenever $G_5 > 0$, meaning that the spin rotation affects both the dynamic and static behaviors of the elastic nematics. We can roughly evaluate the inertia effects of internal rotations. Assuming a typical size of molecular cluster involving in determination of director to be equal of $\sim 10^{-4}$ cm, results in evaluation of inertia moment, $I \sim 10^{-8}$ cm$^2$. Than for such a low arbitrary value of $G_5$ as $\sim 0.1$ Pa, the typical frequency of inertia oscillations for (22b) is evaluated as ~10KHz, i.e. it belongs to the ultrasound frequency range.

## 5. Stability conditions, soft deformation modes, and rotational invariance.

Far away from the order-disorder phase transition, a nematic elastic solid can be considered as thermodynamically stable, with the stability conditions requiring the quadratic form in (15) to be positive definite. To establish these conditions, we introduce a special orthogonal coordinate system $\{\hat{\underline{x}}\}$, whose one axis, say $\hat{x}_1$, is directed along the axis of initial director, $\underline{n}_0$. In this coordinate system,

$$\underline{n}_0 = \{1,0,0\}; \quad \underline{n}_0 \underline{n}_0 = \begin{pmatrix} 1 & 0 & 0 \\ 0 & 0 & 0 \\ 0 & 0 & 0 \end{pmatrix}; \quad \underline{\underline{E}} = \hat{\underline{\underline{E}}}; \quad \underline{\underline{\Omega}}^r = \hat{\underline{\underline{\Omega}}}^r. \qquad (23)$$

Here the anti-symmetric tensor $\underline{\underline{\Omega}}^r$ characterizing relative rotations is frame invariant. Then in $\{\hat{\underline{x}}\}$ coordinates, formula (15) for the free energy is represented in the form:

$$\rho \hat{F} = 1/2 G_s s_r^2 + G_{sa} s_r E_{11} + (1/2 G_0 + G_1 + G_2) \hat{E}_{11}^2 + 1/2 G_0 (E_{22}^2 + E_{33}^2 + 2E_{23}^2) + G_5 (\Omega_{23}^r)^2$$
$$+ \sum_{k=2,3} [(G_0 + G_1) \hat{E}_{1k}^2 + 2 G_3 \Omega_{1k}^r \hat{E}_{1k} + (G_4 + G_5)(\Omega_{1k}^r)^2]. \qquad (24)$$

Except for the first three coupled terms, all other terms in (24) including two last quadratic forms for $k = 2$ and $k = 3$, are independent. Therefore the necessary and sufficient stability conditions are:

$$G_s > 0; \; G_0 > 0; \; G_0 + G_1 > 0; \; 1/2 G_0 + G_1 + \hat{G}_2 > 0; \; G_5 > 0;$$
$$(G_0 + G_1)(G_4 + G_5) > G_3^2; \; (G_4 + G_5 > 0). \qquad (25)$$
$$(\hat{G}_2 = G_2 - 1/2 G_{sa}^2 / G_s)$$

Note that according to (25) the material parameters $G_{sa}$, $G_1$, $\hat{G}_2$, $G_3$ and $G_4$ are sign indefinite.

When using (25), one can also establish the conditions of stability in terms of the de Gennes coefficients:

$$\mu_0 > 0;\ \mu_1 > 0;\ \mu_2 > 0;\ 4\alpha_1\mu_2 > \alpha_2^2 \text{ (i.e. } \alpha_1 > 0). \tag{25a}$$

Using now the coordinate system $\{\hat{x}\}$ in equations (18a,b) represents the expressions for symmetric, $\hat{\sigma}_{ij}^s$, and anti-symmetric, $\hat{\sigma}_{ij}^a$, extra stress tensor components as follows:

$$\hat{\sigma}_{11}^s = (G_0 + 2G_1 + 2\hat{G}_2)\hat{E}_{11};\ \sigma_{22}^s = G_0 E_{22};\ \sigma_{33}^s = G_0\hat{E}_{33} \quad (\sum E_{ii} = 0)$$

$$\hat{\sigma}_{12}^s = (G_0 + G_1)\hat{E}_{12} + G_3\Omega_{12}^r;\ \sigma_{13}^s = (G_0 + G_1)\hat{E}_{13} + G_3\Omega_{13}^r;\ \sigma_{23}^s = G_0 E_{23} \tag{26}$$

$$\hat{\sigma}_{12}^a = G_3\hat{E}_{12} + (G_4 + G_5)\Omega_{12}^r;\ \sigma_{13}^a = G_3\hat{E}_{13} + (G_4 + G_5)\Omega_{13}^r;\ \sigma_{23}^a = G_5\Omega_{23}^r.$$

Here the components $\hat{\sigma}_{kj}^a$ are written only for $k < j$. Note that the strain components $\hat{E}_{22}, E_{33}$ and $E_{23}$ do not produce corresponding nematic contributions in stress and in free energy. This is the effect of uniaxial (nematic) anisotropy.

Except for the particular case,

$$\hat{G}_2 = -G_1, \tag{27}$$

the symmetric extra tensor component $\underline{\underline{\sigma}}^s$ is not traceless. The case (27) is, however, of physical significance. As seen from (15) and (26), only in this case the part of pure nematic contributions in the free energy and stress, proportional to the $G_1 + \hat{G}_2$, does not consist the component of elastic strain $\hat{E}_{11}$ directed along the director. It means that this particular case describes in a sense the "maximal anisotropy". It is easy to see that the particular case (27) does not violate the general stability conditions (25).

Until now the material parameters $G_k$ were considered as independent, however, being under the thermodynamic stability constraints (25). If at least one inequality in the stability constraints (25) changes for equality, it represents the *marginal stability* condition, the term being well known from the general stability theory. The marginal stability imposes some particular relations between the parameters $G_k$, which generally can be valid only for specific types of nematic solids. When a marginal stability condition results in nullifying some components of stress tensor (and, correspondingly, their contributions in the free energy), the respective stress/strain/rotational components are called the *soft deformation modes*. In the nematic solids with the soft deformation modes, new effects, as compared to the traditional anisotropic solid mechanics, may happen. When the only soft mode strains are imposed, they theoretically meet no resistance and do not contribute in the free energy. Warner et al (1994) analysed first these unusual deformation modes in nematic solids. The soft deformations present a special case of the soft mode class for anisotropic materials, generally predicted first by Golubovich and Lubensky (1989), who also proposed a physical explanation for this type of mechanical behavior.

In the following analysis we assume the conditions, $G_0 > 0$ and $G_3 \neq 0$, that prevent the above constitutive relations from degeneration. Using this assumption it is clear that the possible soft modes in (26) can be searched from the conditions: (i) $\hat{\sigma}_{1k} = 0$ ($k$ = 2,3) and (ii) $\hat{\sigma}_{11} = 0$. The condition (i) yields:

$$\hat{\Omega}_{1k} = -E_{1k}(G_0 + G_1)/G_3 = -E_{1k}G_3/(G_4 + G_5) \quad (k = 2,3) . \tag{28}$$

Equalities (28) result in:

$$(G_0 + G_1)(G_4 + G_5) = G_3^2 . \tag{29}$$

When comparing (29) to the stability condition $(G_0 + G_1)(G_4 + G_5) > G_3^2$ in (24), it is clear that (29) is the marginal stability condition. It is also easy to see that (28) and (29) nullify the two last quadratic forms in (24). Thus the shearing modes {12}, {2,1} and {13}, {3,1} in the coordinate system $\{\hat{\underline{x}}\}$ are the soft deformation modes. As seen from (26) and (29), the stress resistances of soft deformations in these modes are completely compensated by the respective contributions of internal rotations in these stresses. This is the reason for occurrence of the soft deformation behaviour. It is also easy to see from (26) that for the soft deformation modes presented in the coordinate system $\{\hat{\underline{x}}\}$, the full anti-symmetric part of stress tensor, $\hat{\underline{\underline{\sigma}}}^a = 0$, whenever the Born coefficient $G_5$ vanishes. Since the condition $\hat{\underline{\underline{\sigma}}}^a = 0$ is the tensor property, it means that in the case $G_5 = 0$, the condition (29) for occurrence of the soft modes necessarily results in the symmetry of stress tensor. Note that the case $G_5 = 0$ is common for all existing constitutive models of weak nematic elasticity.

The possibility (ii), $\hat{\sigma}_{11} = 0$, for the second soft mode to occur results in another marginal stability condition,

$$G_0 + 2G_1 + 2\hat{G}_2 = 0 \tag{29a}$$

Unlike (29) the condition (29a) that does not involve the internal rotations, could be considered as a "pure" marginal stability constraint.

The marginal stability conditions (29) and (29a) can be considered as independent. It is worth noticing, however, that while the single condition (29) does not affect (27) of maximal anisotropy, $\hat{G}_2 = -G_1$, the coupled conditions (29) and (29a) are incompatible with (27).

Olmsted (1994) postulated the principle of "*rotational invariance*" in nematic elastomers to justify the necessity of the soft modes to occur. The rotational invariance means a special type of orthogonal transformation with arbitrary plane rotations about the director, under which some special stress components do not change. These non-changeable stress components are called *rotationally invariant*. It might be directly proved that in coordinate system $\{\hat{x}\}$, the rotationally invariant *shear* stress components must be equal to zero, whereas the rotationally invariant longitudinal stress component $\hat{\underline{\underline{\sigma}}}_{11}$ might not be necessarily equal to zero. It means that in the coordinate system $\{\hat{x}\}$, the rotationally invariant shear nematic deformation modes are soft. This postulated rotational invariance simply follows from the fact that due to uniaxial nematic symmetry there is not a preference in choosing the axes $\hat{x}_2$ and $\hat{x}_3$ in the orthogonal coordinate system $\{\hat{x}\}$. This argument is also confirmed by a simple calculation. Indeed, introducing a special transformation of stress, $\underline{\underline{\sigma}}' = \underline{\underline{q}}^T \cdot \underline{\underline{\sigma}} \cdot \underline{\underline{q}}$ with orthogonal matrix $\underline{\underline{q}}$ depending on the rotational angle $\alpha$, one can simply find the possible rotational invariant stress components $\hat{\sigma}_{ik}$ from the conditions: $\forall \alpha : \sigma'_{ik} = \sigma_{ik}$. It is easy to establish that transformed stress components $\sigma'_{22}, \sigma'_{22}$ and $\sigma'_{33}$, can never be rotationally invariant if they are not degenerated. On the other hand, using the direct calculations one can obtain the conditions of rotational invariance as:

$$\sigma'_{11} \equiv \sigma_{11}, \quad \sigma'_{12} \equiv \sigma_{12} \cos\alpha + \sigma'_{13} \sin\alpha = \sigma_{12}, \quad \sigma'_{13} \equiv -\sigma_{12} \sin\alpha + \sigma'_{13} \cos\alpha = \sigma_{13}. \quad (30)$$

The same conditions are valid for the transponent components of the tensor $\hat{\underline{\underline{\sigma}}}'$. Using the above conditions of rotational symmetry, $\forall \alpha : \sigma'_{ik} = \sigma_{ik}$, relations (30) yield: $\sigma'_{12} = \sigma_{21} = 0$ and $\sigma'_{13} = \sigma_{31} = 0$, whereas the rotationally invariant longitudinal stress component $\hat{\underline{\underline{\sigma}}}_{11}$ is not necessarily equal to zero. On the contrary, if the deformation modes $\{1,2\}, \{2,1\}$ and $\{1,3\}, \{3,1\}$ are soft, i.e. $\sigma'_{12} = \sigma_{21} = \sigma'_{13} = \sigma_{31} = 0$, expressions (30) yield: $\sigma'_{12} = \sigma'_{21} = \sigma'_{13} = \sigma'_{31} = 0$, which means the rotational invariance of these shear stress components.

The results obtained above can be formulated as the extended Olmstead *Theorem* (I):

1. The shear (in the coordinate system $\{\hat{\underline{x}}\}$) soft deformational modes may exist if and only if they are rotationally invariant.
2. In the case of deformational free energy density (15), the existence of the soft deformation modes guarantied only if the marginal stability condition (29) is fulfilled.
3. When the soft shearing deformation modes exist and the Born term is negligible ($G_5 = 0$) the stress tensor is symmetric.

The most important consequence of the theorem I is that in the general case of free energy (15), the translation invariance does not necessitate the existence of even shearing (in the coordinate system $\{\hat{\underline{x}}\}$) soft deformation modes. They occur only when the marginal stability condition (29), which relates several material constants, is satisfied. In this regard, the attempt to necessitate the existing of soft deformation modes by using the "broken symmetry" argument in nematics (e.g. see Lubensky et al (2002)) is also invalid. This is because in this case, the broken symmetry means nothing more than the change of the isotropic symmetry after the disorder-order phase transition for uniaxial (nematic) symmetry. But as shown few lines above, the occurrence of the nematic state does not necessitate existing of the soft deformation modes.

It has been shown in the Section 4, that the effect of internal spin is negligible for nematic elastomers, and the effect of the scalar order parameters is also insignificant for equilibrium (static) situations. Therefore the assertion 3 of the theorem I present a good opportunity to reduce the complicated general approach to significantly more simple formulation. This will be done in the next Section

The soft deformation modes analysed above are an idealization of real processes that might have the semi-soft deformation behavior observed for nematic elastomers (e.g. see Warner (1999) and references there). It is clear that the semi-soft deformation modes cannot be described within the frames of the present deformation theory. One possible way to describe them is to involve in the theory the Frank elasticity effects.

## 6. Reduced formulation and soft deformation modes

In many cases, the non-symmetry of stresses can be neglected. This leads to a highly simplified version of the theory. As seen from (12), it happens when the couple stress tensor $\underline{\underline{\mu}}$, inertia effects of internal rotations and body moment can be neglected either separately or as a sum. The latter case, when they might be in an intrinsic quasi-equilibrium, happens if the Born term is negligible and the soft or semi-soft deformational modes do exist. Unlike suspensions, the inertia effects of internal rotations can be neglected for molecular nematic solids and liquids in a typical mechanical region of frequencies. The stress symmetry assumption might be verified within the continuum approach only when comparing the predictions of the simplified theory with experiments. It should also be noted that the stress symmetry assumption works very well for the common (non-nematic), isotropic and anisotropic elastic solids with small and large deformations.

Using the condition $\underline{\underline{\sigma}}^a = 0$ in (18b) expresses the anti-symmetric tensor of relative rotations $\underline{\underline{\Omega}}^r$ via strain tensor $\underline{\underline{E}}$ and initial value of director $\underline{n}_0$ as:

$$\underline{\underline{\Omega}}^r = \Lambda(\underline{\underline{E}} \cdot \underline{n}_0 \underline{n}_0 - \underline{n}_0 \underline{n}_0 \cdot \underline{\underline{E}}); \quad \Lambda = G_3/(G_4 + G_5). \tag{31}$$

Relation (31) has a sense, since due to the stability conditions (18), $G_4 + G_5 > 0$. Using (31) it is easy to show that $\underline{\underline{\Omega}}^r \cdot \underline{n}_0 = 0$. It means that in the reduced formulation, the spin of internal rotations does not vanish but is equal to the spin of body rotation, i.e. $\underline{\underline{\Omega}}^I_{\parallel} = \underline{\underline{\Omega}}_{\parallel}$, and in (32), $\underline{\underline{\Omega}}^r = \underline{\underline{\Omega}}^r_{\perp}$. Using (6) one can also obtain:

$$\underline{n} \approx (\underline{\underline{\delta}} - \underline{\underline{\Omega}}) \cdot \underline{n}_0 + \Lambda[\underline{\underline{E}} \cdot \underline{n}_0 - \underline{n}_0 tr(\underline{\underline{E}} \cdot \underline{n}_0 \underline{n}_0)]. \tag{32}$$

Substituting (32) into (15) and (18a) yields the expressions for the reduced free energy $F^r$ and reduced symmetric extra stress $\underline{\underline{\sigma}}^r$ as follows:

$$\rho F^r = \frac{1}{2} G_0^r tr \underline{\underline{E}}^2 + G_1^r tr(\underline{n}_0 \underline{n}_0 \cdot \underline{\underline{E}}^2) + G_2^r tr^2(\underline{n}_0 \underline{n}_0 \cdot \underline{\underline{E}}) \tag{33}$$

$$\underline{\underline{\sigma}}^r = G_0^r \underline{\underline{E}} + G_1^r(\underline{n}_0 \underline{n}_0 \cdot \underline{\underline{E}} + \underline{\underline{E}} \cdot \underline{n}_0 \underline{n}_0) + 2G_2^r \underline{n}_0 \underline{n}_0 tr(\underline{\underline{E}} \cdot \underline{n}_0 \underline{n}_0). \tag{34}$$

Here

$$G_0^r = G_0; \quad G_1^r = G_1 - G_3 \Lambda; \quad G_2^r = \hat{G}_2 + G_3 \Lambda, \tag{35}$$

and $\hat{G}_2$ is defined in (19).

Relations (31)-(35) form the closed set of constitutive equations. It is remarkable that Equations (33) and (34) predict the potential relation for reduced stress, $\underline{\underline{\sigma}}^r = \rho \partial F^r / \partial \underline{\underline{E}}$. If the strain tensor $\underline{\underline{E}}$ is known, the internal rotations and actual value of director in deformed state are easily established with the use of Equations (31) and (32), involving no additional parameters than those presented in (35). Here the body rotations are established using the compatibility equation. Note that as in the previous Section, we can substitute within the same precision the initial value of director $\underline{n}_0$ for the actual one $\underline{n}$ in all the equations of this Section except for (32), which will be reduced to:

$$\underline{n}_0 \approx (\underline{\underline{\delta}} + \underline{\underline{\Omega}}) \cdot \underline{n} + \Lambda[\underline{n}\ tr(\underline{\underline{E}} \cdot \underline{nn}) - \underline{\underline{E}} \cdot \underline{n}]. \tag{32a}$$

In order to establish in the reduced approach the stability conditions for free energy and conditions for existence of the soft deformation modes, we use the expression for the reduced free energy (33) in the coordinate system $\{\hat{x}\}$:

$$\rho \hat{F}^r = (1/2 G_0^r + G_1^r + G_2^r)\hat{E}_{11}^2 + (G_0^r + G_1^r)(\hat{E}_{12}^2 + \hat{E}_{13}^2) + 1/2 G_0^r(\hat{E}_{22}^2 + \hat{E}_{33}^2 + 2\hat{E}_{23}^2) \tag{36}$$

Since all three terms in (36) are independent, the stability conditions are:

$$G_0^r > 0; \quad G_0^r + G_1^r > 0; \quad 1/2 G_0^r + G_1^r + G_2^r > 0. \tag{37}$$

Using (35), it is easy to see that the inequalities (35) follow from the general stability conditions (25). In the coordinate system $\{\hat{x}\}$, the components of extra stress tensor are represented as:

$$\begin{aligned}\hat{\sigma}_{11}^r &= (G_0^r + 2G_1^r + 2G_2^r)\hat{E}_{11}; & \sigma_{22}^r &= G_0^r E_{22}; & \sigma_{33}^r &= G_0^r E_{33}; \\ \hat{\sigma}_{12}^r &= (G_0^r + G_1^r)\hat{E}_{12}; & \sigma_{13}^r &= (G_0^r + G_1^r)E_{13}; & \sigma_{23}^r &= G_0^r E_{23}\end{aligned} \tag{38}$$

It is seen that in the case of maximal anisotropy, when $G_1^r + G_2^r = 0$, the contribution of $\hat{E}_{11}$ in both the reduced free energy and reduced stress will be only due to the isotropic terms, proportional to $G_0^r$. As follows from the second inequality in (38), the sign of $G_1^r$ is still arbitrary in this case too.

Evidently, the soft (shearing) deformation modes also occur in the reduced formulation for elastic nematics, whose material parameters $G_k^r$ from (36) satisfy the marginal stability condition (compare to the second stability condition in (35)):

$$G_0^r + G_1^r = 0. \tag{39}$$

Then Equations (36) and (38) show that applying the shear strains $\hat{E}_{12}$ or $\hat{E}_{13}$ (or both) to a nematic elastic body, initially oriented in the direction $x_1$, causes no occurrence of either corresponding stresses $\hat{\sigma}_{12}^r$ or/and $\hat{\sigma}_{13}^r$, or contribution in the reduced free energy $\hat{F}^r$. Substituting (35) with definition of $\Lambda$ from (31) into (39) reduces (39) to the general condition (29).

Note that in the nematic "maximal anisotropic" case, when $G_1^r + G_2^r = 0$, the condition (39) in the reduced potential, allows us to express all the coefficients in formulae (33) and (34) via only $G_0^r$:

$$G_1^r = -G_0^r, \quad G_2^r = G_0^r \quad (G_0^r > 0). \tag{40}$$

When along with marginal stability constraint (39), another (extensional) marginal stability constraint,

$$G_0^r + 2G_1^r + 2G_2^r = 0, \tag{39a}$$

is used, then instead of (40) one obtains:

$$G_1^r = -G_0^r, \quad G_2^r = 1/2 G_0^r \, (G_0^r > 0). \tag{40a}$$

It is seen from (36) that this case corresponds to the minimum of free energy under given values of strains.

Thus in the case (40) when $G_5 = 0$ and shearing soft modes with the maximum symmetry condition are presented, Equations (33) and (34) take the form:

$$\rho F^r / G_0^r = 1/2 \, tr \underline{\underline{E}}^2 - tr(\underline{n}_0 \underline{n}_0 \cdot \underline{\underline{E}}^2) + tr^2(\underline{n}_0 \underline{n}_0 \cdot \underline{\underline{E}}) \tag{33a}$$

$$\underline{\underline{\sigma}}^r / G_0^r = \underline{\underline{E}} - \underline{n}_0 \underline{n}_0 \cdot \underline{\underline{E}} - \underline{\underline{E}} \cdot \underline{n}_0 \underline{n}_0 + 2\underline{n}_0 \underline{n}_0 tr(\underline{\underline{E}} \cdot \underline{n}_0 \underline{n}_0). \tag{34a}$$

In the case (40a) when $G_5 = 0$ and shearing and extensional soft modes are presented, Equations (33) and (34) take the form:

$$\rho F^r / G_0^r = 1/2 \, tr \underline{\underline{E}}^2 - tr(\underline{n}_0 \underline{n}_0 \cdot \underline{\underline{E}}^2) + 1/2 tr^2(\underline{n}_0 \underline{n}_0 \cdot \underline{\underline{E}}) \tag{33b}$$

$$\underline{\underline{\sigma}}^r / G_0^r = \underline{\underline{E}} - \underline{n}_0 \underline{n}_0 \cdot \underline{\underline{E}} - \underline{\underline{E}} \cdot \underline{n}_0 \underline{n}_0 + \underline{n}_0 \underline{n}_0 tr(\underline{\underline{E}} \cdot \underline{n}_0 \underline{n}_0). \tag{34b}$$

**7. Example: infinitesimal Warner potential**

Warner *et al* (1993) using entropy concept derived the following free energy expression for nematic elastomers:

$$2\rho F^w / G = tr(\underline{\underline{l}}_0 \cdot \underline{\underline{F}} \cdot \underline{\underline{l}}^{-1} \cdot \underline{\underline{F}}^T) \tag{41}$$

Here $\underline{\underline{l}}_0$ and $\underline{\underline{l}}$ are the tensors characterizing anisotropy in initial (non-deformed) and actual (deformed) states, and $\underline{\underline{F}}$ is the strain gradient tensor. In (41),

$$\underline{\underline{l}}(\underline{n}) = l_\perp \underline{\underline{\delta}} + (\Delta l)\underline{n}\underline{n}; \quad \underline{\underline{l}}^{-1}(\underline{n}) = 1/l_\perp \underline{\underline{\delta}} - \Delta l /(l_\perp l_\parallel)\underline{n}\underline{n};$$
$$\underline{\underline{l}}_0 = l_\perp^0 \underline{\underline{\delta}} + (\Delta l^0)\underline{n}_0 \underline{n}_0; \quad \Delta l^0 = l_\parallel^0 - l_\perp^0; \quad \Delta l = l_\parallel - l_\perp. \tag{42}$$

The parameters of macromolecular chain anisotropy $l_\parallel$ and $l_\perp$ depend on the nematic scalar order parameter $s$. Explicit form of these dependences is different for different models of nematic polymers. Unlike the common assumption that the direction of preferred orientation changes only due to the action of external fields, we assume here that in a mechanical field, parameter $s$ can also be slightly changed as described in Section 4. It yields that $l_\parallel = l_\parallel^0 + O(s_r)$ and $l_\perp = l_\perp^0 + O(s_r)$, where $s_r = s - s_0$. Thus taking into account the formulae (15) and (18c), we can extend the Olmsted (1994) calculations of the Warner potential for weakly elastic case to:

$$\frac{\rho F^w}{G_o} = -1/2\alpha tr^2(\underline{n}\underline{n} \cdot \underline{\underline{E}}) + 1/2tr(\underline{\underline{E}}^2) + \frac{(\Delta l)^2}{4 l_\perp l_\parallel}\{tr(\underline{n}\underline{n} \cdot \underline{\underline{E}}^2) - tr^2(\underline{n}\underline{n} \cdot \underline{\underline{E}}) - tr[\underline{n}\underline{n} \cdot (\underline{\underline{\Omega}}^r)^2]\}$$
$$- \frac{\Delta l^2}{2 l_\perp l_\parallel} tr[\underline{n}\underline{n} \cdot (\underline{\underline{E}} \cdot \underline{\underline{\Omega}}^r)] \; ; \; (\Delta l)^2 = (l_\parallel - l_\perp)^2; \; \Delta l^2 = l_\parallel^2 - l_\perp^2; \; \alpha = \frac{2G_{sa}^2}{G_s G_0} > 0. \tag{43}$$

To simplify the notations we omitted the "zero" indexes in $\underline{n}$, $l_\parallel$, and $l_\perp$.

Comparing (43) with (15), where $s_r$ has been substituted using (18c), one can notice that (43) presents a very particular version of the general potential (15), with the following correspondence between material parameters in (15) and in (43):

$$G_1 = -G_2 = G_4 = G_0 \frac{(l_\parallel - l_\perp)^2}{l_\parallel l_\perp}; \quad G_3 = G_0 \frac{l_\parallel^2 - l_\perp^2}{l_\parallel l_\perp}; \quad G_5 = 0, \alpha = \frac{G_{sa}^2}{G_s G} > 0. \tag{44}$$

Equations (18) are valid for extra stress in the case of potential (43) with specification (44). It is easy to see that these parameters, along with parameter $\alpha$, satisfy the thermodynamic stability conditions (25).

The most striking fact regarding the Warner potential (43) is that its parameters shown in (44) identically satisfy the general marginal condition (29). It means that the Warner potential (43) always predicts the existence of soft deformation modes. Because the Born term is absent ($G_5 = 0$), the Warner potential also predicts that $\underline{\underline{\sigma}}^a = 0$. It means that the expressions for the Warner potential and related stress tensor are always equivalently reducible in the sense of the previous Section.

We now briefly describe the equivalent formulation for the Warner potential and respective stress, using the reduction procedure of the previous Section. According to (44), the parameter $\Lambda$ in (31), (32) and the parameters $G_k^r$ ($k = 0,1,2$) in (24) are of the form:

$$\Lambda = \Delta l^2 / (\Delta l)^2 \,; \quad G_0 = -G_1^r = G_2^r. \tag{45}$$

Due to (45) the formulae for reduced Warner potential $F_w^r$ and the extra stress $\underline{\underline{\sigma}}_w^r$ are:

$$\rho F_w^r / G_0 = 1/2 \, tr\underline{\underline{E}}^2 - tr(\underline{nn} \cdot \underline{\underline{E}}^2) + (1-\alpha) tr^2(\underline{nn} \cdot \underline{\underline{E}})], \tag{46}$$

$$\underline{\underline{\sigma}}_w^r / G_0 = \underline{\underline{E}} - \underline{nn} \cdot \underline{\underline{E}} - \underline{\underline{E}} \cdot \underline{nn} + 2(1-\alpha)\underline{nn}\,tr(\underline{\underline{E}} \cdot \underline{nn})]. \tag{47}$$

Remarkable that in particular cases, $\alpha = 0$ and $\alpha = 1/2$, the relations (46), (47) following from detailed Warner model, coincide with Equations (33a), (34a) and (33b), (34b), respectively, which follow from general expression (15) for the free energy function. Moreover, Equations (46), (47) cover the intermediate cases, too.

## 8. Conclusions

1. Using the general expression (15) for the free energy function, the paper develops a general theory of weak (linear) elasticity for nematic solids. The need for such a general theory, even in linear limit, is motivated by its application to new emerging materials, such as nanocomposites and bio-materials like soft tissue, whose possible nematic behavior may not satisfy the main assumptions valid for nematic elastomers. There were two main objectives of our research: (i) better understanding of the general conditions for occurrence of the soft deformation modes, and (ii) developing a simplified formulation that still captures the basic features of general theory and could be used in more complicated non-equilibrium cases.

2. Using the general free energy function (15), it has been proven that neither rotational invariance nor broken symmetry arguments could justify the necessity for soft deformation modes to occur.

3. Although the effect of internal spin rotations and was found insignificant in the case of elastic elastomers, the effect of change in scalar order parameter might be significant. Both these effects might be significant for theoretical description of equilibrium and non-equilibrium effects in nematic solid and liquids systems, such as polymer suspensions and composites.

4. A simplified, reduced formulation of nematic effects in elastic solids has been developed, which neglects the effect of stress non-symmetry in the general case when there is no soft deformation behavior.

5. In the particular case of soft deformation behavior when the symmetry of stress is valid, the reduced approach being exact, reduces to a one-parametric, weak nematic elastic theory.

It should be mentioned that the weak Warner potential supported by existing molecular models, satisfies the marginal stability constraint and therefore predicts the existence of only soft nematic modes. Therefore it completely describes the situation in 5. Additionally, the only soft modes have been found till now in experiments for a class of nematic elastomers.